\documentclass[runningheads]{llncs}
\usepackage[T1]{fontenc}
\usepackage{graphicx}
\usepackage{xcolor}

\usepackage{booktabs}
\usepackage{multirow}
\usepackage{array}
\usepackage{amssymb}
\usepackage{hyperref}
\usepackage{tcolorbox}
\tcbuselibrary{breakable}
\usepackage{xcolor}
\usepackage{colortbl}
\usepackage{subcaption}
\usepackage{amsmath} 
\usepackage{enumitem} 
\usepackage{float}
\usepackage[section]{placeins}

\usepackage{listings}

\newtcolorbox{promptbox}[1]{
  colback=gray!5,
  colframe=gray!45,
  title=#1,
  fonttitle=\bfseries,
  coltitle=black,
  arc=1mm,
  boxrule=0.4pt,
  left=6pt,
  right=6pt,
  top=6pt,
  bottom=6pt,
  breakable
}

\definecolor{contexthl}{RGB}{255,250,205} 
\definecolor{structhl}{RGB}{173,216,230}  

\newcommand{\myparagraph}[1]{\paragraph{\textbf{#1}}\mbox{}\\}

\begin{document}
\title{Scalable Generation and Validation of Isomorphic Physics Problems with GenAI}
\titlerunning{Scalable Generation and Validation of Isomorphic Physics Problems}
%
\author{
Naiming Liu\inst{1}
\and Leo Murch\inst{2}
\and Charles Alliotts\inst{2}
\and Spencer Moore\inst{2}
\and Tong Wan\inst{2}
\and Shashank Sonkar\inst{2}
\and Richard Baraniuk\inst{1}
\and Zhongzhou Chen\inst{2}\thanks{Corresponding author: \email{zhongzhou.chen@ucf.edu}}
}
\authorrunning{Liu et al.}
%
\institute{
Rice University
\and
University of Central Florida
}
\maketitle              
\begin{abstract}

As traditional physics exams face multiple growing challenges today, a promising alternative is to create exams based on large isomorphic problem banks, which enable asynchronous, multi-attempt exams in which students practice from openly available problem banks and receive exam problems that assess the same concepts through varied surface features. However, creating such banks at scale and validating their consistency in difficulty is resource-intensive. In this paper, we first present a human-in-the-loop GenAI framework for generating isomorphic physics problem banks using prompt chaining and tool use. The framework distinguishes between structural variations and contextual variations, allowing instructors to control solution structure while varying problem contexts. We then evaluate the difficulty of generated banks using both student response data from 20 banks deployed in exams, and retrospective simulations with 17 open-source language models (LMs). Results show that 75\% of deployed banks achieve statistically homogeneous difficulty, as measured by actual student response data. LM simulation classifies 70\% of evaluated banks as homogeneous, and shows moderate alignment with student difficulty patterns for some banks. Two case studies further show that LM responses can identify potential causes of unintended difficulty, including ambiguous wording, overlooked implicit verbal constraints, and configurations that invite common physics reasoning errors. These findings suggest that GenAI can make large-scale isomorphic physics problem generation practical, while LM simulation can complement human review as a low-cost pre-deployment screening tool.

\end{abstract}

\section{Introduction}
Assessment is a fundamental component of STEM education, yet conventional reliance on single-attempt, synchronous exams presents growing challenges \cite{muldoon2012,zilles2019}. In traditional exams, synchronous scheduling plays a central role in maintaining item security and discouraging academic dishonesty. However, requiring all students to sit for exams simultaneously creates a high-stress environment and significant accessibility barriers \cite{zilles2019}, particularly for students with non-traditional schedules or exceptional circumstances. For instructors and researchers, the need to keep test items secure also limits the use of common assessments, restricting comparisons of student learning outcomes across different instructional settings. At the same time, the ability of synchronous exams to uphold item security is increasingly undermined by online resource-sharing platforms and generative AI (GenAI) \cite{Flugge2024,Francis2025,Newton2025}. As a result, researchers have begun exploring asynchronous alternatives that rely on item randomization to provide more flexible scheduling while discouraging rote memorization and collaborative cheating \cite{Frederick2024PERC,Silva2020SIGCSE,Sud2019ASEE,Sullivan2016,Liu2025Frontiers}.

Previous research has proposed that asynchronous, multi-attempt assessments can be supported by large isomorphic problem banks \cite{Liu2025Frontiers,Frederick2024PERC}. In this study, we define isomorphic problems as items that assess the same underlying concepts and solution structures while varying surface features and contexts. For example, the two problems shown in Table~\ref{tab:example} are isomorphic under this definition. Compared with conventional parameterized questions that typically vary only numerical values, isomorphic problems allow for richer variation across diverse contexts while maintaining consistent difficulty \cite{millar2021repeat}. Prior work has shown that when exam problems are randomly drawn from a sufficiently large problem bank, the bank can be made openly available for practice before an exam without compromising its diagnostic effectiveness \cite{Liu2025Frontiers}. Because the volume and diversity of such banks make rote memorization a cognitively inefficient strategy, they can encourage students to develop more transferable knowledge. If each exam problem is drawn from an isomorphic bank that is already available for practice, students no longer need to take the exam at the same time and can be allowed multiple attempts \cite{Frederick2024PERC}. This novel type of exam can increase accessibility and overcome the challenges of conventional exams without sacrificing assessment validity.

\begin{table}[t!]
\centering
\begin{tcolorbox}[colback=blue!5!white, colframe=blue!75!black, title=Problem Variant 1]
A \textcolor{purple}{traveler} is dragging a heavy \textcolor{purple}{backpack} across an \textcolor{purple}{asphalt road} at a constant speed. 
The coefficient of kinetic friction between the \textcolor{purple}{backpack} and the ground is \textbf{\textcolor{blue}{0.73}}. 
The person pulls the \textcolor{purple}{backpack} at an angle of \textbf{\textcolor{blue}{$37.92^\circ$}} upward with respect to the horizontal. 
The force applied by the \textcolor{purple}{traveler} is \textbf{\textcolor{blue}{59.81 N}}. 
The acceleration due to gravity is $9.81 \text{ m/s}^2$. 

\textbf{Calculate the \underline{mass} of the \textcolor{purple}{backpack} in kilograms. Round your answers to two decimal places.}
\end{tcolorbox}

\begin{tcolorbox}[colback=green!10!white, colframe=green!50!black, title=Problem Variant 2]
A \textcolor{purple}{dog} is pulling a small \textcolor{purple}{sled} across an \textcolor{purple}{icy path} at a steady speed. 
The coefficient of kinetic friction between the \textcolor{purple}{sled} and the ice is \textbf{\textcolor{blue}{0.59}}. 
The \textcolor{purple}{sled} has a mass of \textbf{\textcolor{blue}{10.21 kg}}. 
The \textcolor{purple}{dog} pulls the \textcolor{purple}{sled} using a rope, which forms an angle of \textbf{\textcolor{blue}{$31.21^\circ$}} with respect to the horizon. 
The acceleration due to gravity is $9.81 \text{ m/s}^2$. 

\textbf{Determine the \underline{force} exerted by the \textcolor{purple}{dog} in Newtons. Round your answers to two decimal places.}
\end{tcolorbox}
\caption{\textbf{Examples of Isomorphic Problems}. \textcolor{purple}{Purple} denotes contextual variations, \textcolor{blue}{blue} denotes structural variations, and \underline{underlining} marks the unknown variable.}
\label{tab:example}
\end{table}

However, creating and evaluating large isomorphic problem banks at scale still presents several major bottlenecks. First, manually generating the large number of high-quality isomorphic variants needed to populate these banks is prohibitively resource-intensive. Second, even though the practical definition of ``isomorphism" helps maintain structural uniformity, it does not guarantee psychometric uniformity across problem variants. Evaluating and ensuring consistency of difficulty across isomorphic problems is challenging for several reasons. On one hand, each problem in a bank is administered to only a small fraction of students on a given exam, making it difficult to validate the homogeneity of item difficulty using traditional student response data. On the other hand, for conventional multiple-choice or numerical-response problems, response data alone provides little insight into why some isomorphic variants are more difficult than others. Finally, the quality of generated problem banks should ideally be evaluated before they are deployed to students, but such pre-deployment validation is difficult to achieve in traditional educational settings.

We address these challenges by developing a GenAI-based framework for the scalable generation of isomorphic problem banks in introductory Newtonian mechanics. The framework follows a human-in-the-loop workflow built around \textbf{prompt chaining} and \textbf{tool use}, and can be implemented with commercially available AI services such as ChatGPT~\cite{achiam2023gpt}. We evaluate the quality of the generated problem banks through two approaches: a post-deployment analysis of student performance across three midterm exams in a class of more than 220 students and an exploratory pre-deployment analysis using 17 open-source Language Models (LMs) as simulated students. More specifically, we seek to answer the following three research questions:

\begin{description}
    \item[RQ1:] For isomorphic problems generated using the proposed framework, to what extent are the difficulties of individual items homogeneous across a bank?

    \item[RQ2:] To what extent does the performance of open-source LMs on isomorphic problem banks align with actual student performance?

    \item[RQ3:] Can responses from open-source LMs help identify the likely causes of significant heterogeneity in difficulty among generated problems?
\end{description}

To answer these questions, we evaluate 20 deployed problem banks using student performance data, and 20 problem banks using 17 open-source LMs ranging from 0.6B to 32B parameters. The student-based analysis measures the homogeneity of difficulty within each bank, whereas the LM-based analysis compares model performance patterns with student outcomes and examines whether model errors can help diagnose problematic variants. Overall, our results suggest that GenAI can make large-scale isomorphic problem generation practical, and that LM-based evaluation can serve as a useful pre-deployment screening tool for flagging potentially problematic variants before classroom deployment.

The main contributions of this paper are as follows:

\begin{enumerate}
    \item \textbf{GenAI-Based Problem Generation Framework}: We present a human-in-the-loop framework that uses prompt chaining and tool use to generate isomorphic physics problems while maintaining explicit control over structural variations and enabling diverse contextual variations.
    \item \textbf{ESTELA-Physics Resource}: We introduce ESTELA-Physics, a dataset of 29 isomorphic physics problem banks across 12 topics in introductory Physics I, providing a resource for studying scalable isomorphic problem generation and validation.
    \item \textbf{Empirical Validation Study}: We provide a combined student-based and LM-based evaluation of generated problem banks, showing how open-source LMs can complement classroom data when screening for difficulty heterogeneity and diagnosing problematic variants.
\end{enumerate}

The remainder of the paper is organized as follows. We first review related work on automated question generation and LM-based difficulty estimation. We then present the isomorphic problem generation framework, followed by the validation design used to evaluate generated banks. Next, we describe the ESTELA-Physics dataset and report the main results comparing student and LM performance. We also show two case studies diagnosing problem banks identified as heterogeneous based on student performance data. 
Finally, we discuss the implications of our findings, limitations of the current study, and directions for future work.

\section{Related Works}
\subsection{Automated Question Generation}

There has been significant progress in developing Automated Question Generation (AQG) and Automated Item Generation (AIG) technologies in education over the past decade, aiming to reduce the time and cost of item creation while increasing the availability of questions for both assessment and practice \cite{kurdi2020systematic}. Early AQG and AIG approaches primarily rely on hard-coded, template-based methods, which are often time-consuming to develop and require domain-specific programming \cite{kosh2019cost}.
Recent research increasingly explores large language models (LLMs) as a more flexible alternative to fully hand-coded generation pipelines. For example, Dijkstra et al. fine-tune GPT-3 to generate reading-comprehension multiple-choice questions with correct answers and plausible distractors \cite{Dijkstra2022ReadingCQ}. Jiao et al. propose a controllable generation approach for math word problems with specified difficulty levels \cite{Jiao2023AutomaticEQ}. Chan et al. use prompting strategies with GPT-3.5 and GPT-4 to generate assessment items across STEM subjects, including chemistry, physics, and mathematics \cite{chan2025automatic}.

However, evidence from physics question generation suggests that LLM-based AQG remains sensitive to prompting design. Omopekunola and Kardanova \cite{omopekunola2024automatic} evaluate the use of simple prompts to generate multiple-choice items aligned with Bloom's taxonomy. They find that model effectiveness varies substantially across prompting methods and that LLMs struggle to generate items at higher cognitive levels, such as ``Application,'' without highly specific instructional scaffolding. This instability suggests that single-shot prompting is insufficient for complex domain-specific item-generation tasks and motivates the use of multi-prompt strategies, such as the prompt-chaining approach used in the current study.

Most of these studies focus on generating new questions within a given content area, format, or cognitive level. However, as pointed out by Maity et al. \cite{maity2025can}, LLMs often struggle to maintain the intended cognitive level under general prompting conditions and frequently default to factual-recall questions. Furthermore, generating homogeneous and effective item variants remains a major challenge for LLM-based AQG systems \cite{jmir2026anesthesiology}.

Fewer studies explore the generation of isomorphic problem variants that assess the same underlying concepts and principles while differing in surface features. This line of work traditionally relies on template-based methods combined with structured constraints that explicitly control how variants differ. For instance, Arendasy et al. \cite{arendasy2007using} generate algebra word problems by combining required and optional sentences and manipulating the number of equations. Norberg et al. \cite{norberg2023rewriting} use GPT-4 to rewrite the explanations for math word problems, leveraging the model's Python code-generation capabilities to ensure solution correctness. However, recent findings indicate that simply reframing standard problems into scenario-based versions with extraneous contextual details can substantially degrade LLM performance, even when the underlying solution structure appears unchanged \cite{deepquestion2025arxiv}.

One challenge in using LLMs to create isomorphic problems is that they struggle to consistently follow instructions on some dimensions while still producing meaningful diversity on others. Consequently, recent work increasingly favors a ``human-in-the-loop" workflow, in which AQG systems act as drafting agents that generate candidate items for expert review rather than functioning as fully autonomous pipelines \cite{researchgate2025lmaig}. A second challenge is diagram generation. Many physics problems require precise visual representations, but current LLMs still have limited capability to precisely control spatial relationships and visual details with the accuracy needed for assessment. Overall, prior work suggests that LLMs can reduce the cost of question generation, but controlled generation of homogeneous isomorphic variants still requires structured constraints and human oversight.

\subsection{Estimating Question Difficulty using LLMs}

Estimating item difficulty is central to educational assessment because it affects test validity, score interpretation, adaptive testing, and fairness. Traditionally, difficulty is estimated after deployment from student response data, often using classical test theory (CTT) or item response theory (IRT)~\cite{hambleton1993comparison}. However, these methods require sufficient response data and are difficult to apply to newly generated items before classroom use. This ``cold-start'' problem motivates automated item difficulty prediction, which estimates difficulty from item text, metadata, linguistic features, or model-based representations.

Early automated approaches often treat difficulty prediction as a supervised modeling problem, using item features such as readability, syntactic complexity, semantic representations, and transformer-based embeddings. Recent reviews show that these text-based methods can predict empirical item difficulty with moderate to strong accuracy in some settings~\cite{peters2025text}. Related shared-task work further expands this line of research to predicting both item difficulty and response time for multiple-choice questions~\cite{yaneva2024findings}. Several participating systems incorporate LLM-derived features, such as zero-shot LLM responses~\cite{rogoz2024unibucllm} or combinations of linguistic features and LLM prompting~\cite{veeramani2024large}.

Another line of work uses LLMs more directly as simulated students. Rather than only extracting textual features from items, these approaches prompt or align LMs to produce responses that approximate the behavior of students with different knowledge states or ability levels. Lu et al. \cite{lu2024generative} propose \textit{Generative Students}, which uses knowledge-component-based student profiles to simulate responses to multiple-choice questions and identify difficult items that overlap with those identified from real student data. Similarly, Park et al. \cite{park2024large} propose LLaSA, which represents students at different ability levels using LLMs and estimates question difficulty with or without student response records.

Recent studies also align simulated students more explicitly with psychometric models. Scarlatos et al.~\cite{scarlatos2025smart} propose SMART, which aligns simulated students with IRT-based ability levels by generating simulated responses, scoring them with an LLM-based model, and fitting an IRT model to estimate item difficulty. This approach is relevant to pre-deployment validation because it addresses the challenge of estimating difficulty for unseen items without real student responses. However, LLMs should not be treated as perfect student proxies. Liu et al.~\cite{liu2025llms} find moderate correlations between LLM-assigned probabilities and actual student response distributions on multiple-choice questions, especially for distractor choices, suggesting that LLMs can capture some student-like error patterns but do not fully align with human reasoning.

Overall, these studies suggest that LLMs can provide useful, low-cost signals for estimating item difficulty, identifying confusing distractors, and supporting pre-deployment item review. However, most existing work focuses on independent items rather than on validating difficulty homogeneity within isomorphic problem banks. This distinction is important because isomorphic banks are intended to contain variants that assess the same concepts and solution structures; if surface-level changes in context, wording, numerical values, or spatial configuration make some variants easier or harder than others, fairness in randomized or asynchronous assessments may be compromised. Our work addresses this gap by using LM simulations to evaluate whether variants within the same bank differ in difficulty, whether those patterns align with student performance, and whether model responses can help explain why certain variants are more difficult than intended. In this way, LM-based simulation serves as a pre-deployment quality-assurance mechanism for scalable isomorphic question generation.

\section{Isomorphic Problem Generation}
\label{sec:framework}
\subsection{AI-assisted Isomorphic Problem Generation}

Our approach leverages prompt chaining and tool use to provide precise control over structural variations while preserving flexibility in contextual generation. \textbf{Prompt chaining} decomposes complex generation tasks into sequential sub-tasks, where outputs from earlier prompts inform subsequent ones and support multi-step reasoning beyond single-prompt capabilities. For isomorphic problem generation, this process helps separate construct-relevant variations (e.g., numerical values, spatial arrangements) from construct-irrelevant variations (e.g., context, wording), so that each can be constrained independently. \textbf{Tool use} enables LLMs to invoke external functions beyond text generation. In our framework, the Python code interpreter tool~\cite{achiam2023gpt} executes LLM-generated scripts in real time to systematically generate constrained variations, validate solution correctness, and, when needed, produce simple diagrams aligned with problem specifications.

\subsection{Problem Generation Framework}

Our framework distinguishes between two types of variation. \textbf{Structural variations} are construct-relevant changes (e.g., numerical values, spatial arrangements, number of objects) that must remain within precise ranges to ensure correctness and appropriate difficulty. These variations are typically generated by prompting LLMs to invoke Python code execution through \textbf{tool use}. \textbf{Contextual variations} modify surface features (e.g., problem context, scenarios, objects) with fewer restrictions but greater creative flexibility. The restrictions may reflect factors such as student reading level, language proficiency, or cultural background.  Based on these distinctions, we developed the following seven-step framework for prompt-chain design:

\begin{enumerate}
    \item Identify a template problem or problem type.
    \item Identify the problem components.    
    \item Define structural and contextual variations with their constraints.
    \item Design a prompt chain to generate variations for individual components. 
    \item Execute and iteratively refine each prompt based on outcomes. 
    \item Combine components into complete problems in the desired format.
    \item Verify the correctness of generated problems using LLMs.
\end{enumerate}

In this paper, we use \textit{\textbf{structural variations}} to refer to construct-relevant changes to the core structure of the problem that must stay within the user-defined range to preserve the correctness and appropriate difficulty of the problem. Common examples include numerical values of variables, spatial arrangement and number of objects (e.g., forces, particles) in the problem. These variations are often created through a combination of direct LLM output and Python code execution. For instance, when generating numerical values for variables, the Python interpreter can enforce strict constraints, such as ensuring that the applied horizontal force is either less than or greater than the maximum static friction, depending on the intended physical situation.

We use \textbf{\textit{contextual variations}} to refer to changes in the surface features of a problem. These variations are subject to less stringent restrictions but require greater contextual creativity from the LLM. Such constraints may also account for pedagogical considerations, including students' reading level, language proficiency, cultural background, and life experiences. 

Structural and contextual variations may also interact with each other. For example, the numerical value of an object's weight (structural variation) should be constrained by the type of object described in the problem context (contextual variation). Managing these interactions is therefore an important consideration in prompt-chain design.

\subsection{Example Problem Bank Creation: Angled Force with Friction}

We illustrate the problem generation framework using problem bank \textit{3-3}, which focuses on angled forces with kinetic friction.

\noindent\textbf{Problem Template:} An object is pushed or pulled by a single angled force across a rough surface at constant velocity. Students are asked to calculate either the mass, force, or coefficient of kinetic friction in different problem versions.

\noindent\textbf{Problem Components:} Each problem consists of a problem body, a set of known and unknown variables, and a solution.

\noindent\textbf{Variations and Constraints:}
\begin{itemize}
    \item \textbf{Contextual Variation:} Realistic scenarios of pushing/pulling objects on rough surfaces with angled forces (e.g., horse pulling sledge on snow, person pushing couch on carpet).
    \item \textbf{Structural Variation:} The main structural variations include: (1) Force direction and nature (upward/downward, pushing/pulling); (2) Known variable values (friction coefficient, force angle, mass, force magnitude); (3) Unknown variable selection.
    \item \textbf{Structural Constraints:} Angle of force 10°–60° from horizontal; mass and force values appropriate to context; horizontal force component balances kinetic friction; force magnitude must be positive; unknown variable chosen from mass, force, or friction coefficient (excluding angle for consistent computational complexity).
\end{itemize}

\noindent\textbf{Prompt Chain Design:} The prompt chain consists of five prompts. For this case, contextual variation is generated before structural variation because the context informs the selection of variable values. Since some of the prompts are quite lengthy, we summarize the goal of each prompt below. The full prompts used to generate this problem bank are presented in Appendix~\ref{app:a}.

\textbf{\textit{Prompt 1}}: Generate 10 problem contexts with brief descriptions (e.g., ``horse pulling sledge on snow, upward").

\textbf{\textit{Prompt 2}}: Generate context-appropriate random variable values that satisfy the structural constraints. Use Python to calculate the required force using $F = \mu mg/(\cos\theta + \mu\sin\theta)$, and verify that it is positive and falls within realistic bounds.

\textbf{\textit{Prompt 3}}: Select different unknown variables across problem versions and compose problem bodies. This prompt is iteratively refined to specify units and significant figures.

\textbf{\textit{Prompt 4}}: Generate step-by-step solutions following specified format guidelines. This prompt is also refined through multiple iterations.

\textbf{\textit{Prompt 5}}: Export the problems and solutions in the predefined data format.

\section{ESTELA-Physics Dataset}
\label{sec:dataset}
\begin{table}[t!]
\centering
\setlength{\tabcolsep}{6pt}
\setlength{\abovecaptionskip}{6pt}
\resizebox{\columnwidth}{!}{
\begin{tabular}{cccccc}
\toprule
\textbf{Topic} & \textbf{ID} & \textbf{{Problem Bank}} & \textbf{\#Items} & \textbf{Type} & \textbf{Img} \\
\midrule

\multirow{2}{*}{Vector and Math} 
  & \textit{0-1} & Unit Conversion & 10 & NUM & \ensuremath{\times} \\
  & \textit{0-2} & Vector Addition and Subtraction & 10 & MCQ & \checkmark \\
\midrule

\multirow{2}{*}{1D Motion}
  & \textit{1-1} & Motion w Constant Acceleration& 10 & MCQ & \checkmark \\
  & \textit{1-2} & Motion w Non-Constant Acceleration & 21 & MCQ & \ensuremath{\times}  \\
\midrule

\multirow{2}{*}{2D Motion}
  & \textit{2-1} & Projectile Flight Time & 26 & NUM & \checkmark \\
  & \textit{2-2} & Partial Projectile Motion & 20 & NUM & \ensuremath{\times} \\
\midrule

\multirow{6}{*}{Forces}
  & \textit{3-1} & Newton's Third Law Pair & 25 & MA & \ensuremath{\times} \\
  & \textit{3-2} & Newton's Third Law Force Pairs & 26 & MA  & \ensuremath{\times} \\
  & \textit{3-3} & Angled Force with Friction & 20 & NUM & \ensuremath{\times} \\
  & \textit{3-4} & Static block on Incline & 48 & NUM & \checkmark \\
  & \textit{3-5} & Static Friction & 15 & MCQ  & \ensuremath{\times} \\
  & \textit{3-6} & Free Body Diagram Forces & 25 & MA  & \ensuremath{\times} \\
\midrule

\multirow{3}{*}{Newton's Law of Motion}
  & \textit{4-1} & Acceleration from Forces & 20 & NUM & \ensuremath{\times} \\
  & \textit{4-2} & Forces and Acceleration & 20 & MCQ & \ensuremath{\times} \\
  & \textit{4-3} & Accelerating Elevator & 28 & CAT  & \ensuremath{\times} \\
\midrule

\multirow{1}{*}{Kinetic Energy and Work}
  & \textit{5-1} & Work by Varying Force & 25 & MCQ & \ensuremath{\times} \\
\midrule

\multirow{2}{*}{Conservation of ME}
  & \textit{6-1} & ME with External Work & 20 & NUM & \ensuremath{\times} \\
  & \textit{6-2} & ME Conservation Conditions & 30 & CAT & \ensuremath{\times} \\
\midrule

\multirow{1}{*}{Momentum and Impulse}
  & \textit{7-1} & Change in Impulse & 35 & NUM & \ensuremath{\times} \\
\midrule

\multirow{2}{*}{Many-particle Systems}
  & \textit{8-1} & 2D Collision & 24 & NUM & \ensuremath{\times} \\
  & \textit{8-2} & Multi-stage Conversation & 24 & CAT & \ensuremath{\times} \\
\midrule

\multirow{4}{*}{Rotational Motion}
  & \textit{9-1} & Composite Moment of Inertia & 26 & MCQ & \checkmark \\
  & \textit{9-2} & Rotational Kinematics & 30 & NUM & \ensuremath{\times} \\
  & \textit{9-3} & Two Torques Rotational Dynamics & 30 & NUM & \ensuremath{\times} \\
  & \textit{9-4} & Torque Balance & 30 & NUM & \checkmark \\
\midrule

\multirow{2}{*}{Angular Momentums}
  & \textit{10-1} & Angular Momentum Conservation & 12 & NUM & \ensuremath{\times} \\
  & \textit{10-2} & Particle Angular Momentum  & 36 & MA & \checkmark \\
\midrule

\multirow{1}{*}{Simple Harmonic Motion}
  & \textit{11-1} & Math Expression for SHM & 20 & MCQ & \ensuremath{\times} \\
\midrule

\textbf{Total} &  &  & \textbf{666} &  &  \\
\bottomrule
\end{tabular}
}
\caption{Dataset statistics for the ESTELA-Physics dataset, organized by topic. Problem types include numerical response (\textbf{NUM}), multiple-choice (\textbf{MCQ}), multiple-answer (\textbf{MA}), and categorization (\textbf{CAT}) questions. The \textbf{Img} column indicates whether a bank contains image-based questions.}
\vspace{-8mm}
\label{tab:dataset_stats_by_chapter}
\end{table}

Following the proposed generation framework, we construct the \textit{ESTELA-Physics} dataset, a collection of isomorphic problem banks for calculus-based University Physics I, with an emphasis on Newtonian mechanics. The dataset contains 29 problem banks spanning 12 topics commonly covered in introductory university physics courses. As summarized in Table~\ref{tab:dataset_stats_by_chapter}, the banks span topics from vectors, kinematics, and forces to energy, momentum, rotational motion, angular momentum, and simple harmonic motion. The complete dataset is publicly available in an open-source Github repository: \url{https://github.com/Zhongzhou/ESTELA-physics-problem-bank}.

Each problem bank is stored in a dedicated folder, and contains at least the following two components:
 \begin{itemize}
     \item A YAML file (.yml or .yaml): A human-readable file format that stores the isomorphic problems and associated metadata, including all prompts used in the generation, and updates to the problem bank after generation.
     \item A QTI file (.zip): A file that can be directly imported into the Canvas learning management system as an Item Bank.
 \end{itemize}
 
Banks with image-based questions also include a ``Figures'' subfolder with the associated images.

Across all banks, the dataset contains 666 individual isomorphic questions, with each bank containing between 10 and 48 items. The dataset includes four assessment formats: numerical response (\textbf{NUM}), multiple-choice (\textbf{MCQ}), multiple-answer (\textbf{MA}), and categorization (\textbf{CAT}). Six banks include image-based questions (as indicated in the \textbf{Img} column). Each problem bank undergoes final validation review by at least one physics faculty member to ensure pedagogical accuracy and technical correctness.

\section{Validation of Isomorphic Problem Banks}
We evaluated the generated isomorphic problem banks using two complementary approaches. First, we used post-deployment student response data from three midterm exams to measure bank difficulty and homogeneity. Second, we conducted a retrospective LM-based evaluation to simulate how pre-deployment screening could flag potentially problematic banks or variants before future classroom use.
  
\subsection{Post-Deployment Classroom Evaluation}

Twenty isomorphic problem banks from the \textit{ESTELA-Physics} dataset were deployed across three midterm exams in a large-enrollment introductory physics course ($N>220$ students) at a public university in the southeastern United States. Each exam comprised seven problems, with each problem randomly drawn from a distinct isomorphic bank. Midterm Exam 3 contained one problem that was not drawn from an isomorphic problem bank, and was excluded from the current analysis dataset. The problem banks deployed on the exams are presented in Table~\ref{tab:dataset_stats_by_chapter} as those with available student data.

All problem banks used on each exam were made available to students for practice one to three weeks prior to the exam. Students were informed that the upcoming exam would contain problems from those banks. Each exam was 45 minutes long, and students could choose between two testing slots offered during the same week. Students were allowed one attempt on each exam. The exams were administered through the Canvas Learning Management System (LMS), and students were required to bring their own computer to complete the exam in a classroom with human proctoring. A lockdown browser was not used; instead, human proctors actively monitored the room to ensure that no other browser windows were open during the exam.

\subsubsection{Collection and Analysis of Student Performance Data.}

Student performance data were collected from the Canvas LMS and de-identified prior to analysis. For the present study, we extracted only binary correctness data for each problem, indicating whether each student response was correct or incorrect.

\subsubsection{Problem Difficulty.}

The difficulty of each isomorphic problem was measured using student accuracy, defined as the proportion of students who answered the problem correctly. Lower accuracy therefore indicates greater item difficulty. The difficulty of a problem bank was calculated as the sample-size-weighted average across all isomorphic problems in that bank.

\subsubsection{Homogeneity of Difficulty.} 

Homogeneity of difficulty within each problem bank was first assessed using the standard deviation (std) of item-level accuracies across isomorphic problems. Lower standard deviation values indicate greater homogeneity. We also applied the extended Fisher's exact test~\cite{upton1992fisher} to assess whether difficulty differed significantly across isomorphic variants within each bank at the $\alpha = 0.05$ level. Problem variants with fewer than five student responses were excluded from this test. Banks were classified as \textit{homogeneous} (p-value $> 0.05$) or \textit{heterogeneous} (p-value $\leq 0.05$) based on the statistical significance of the test results. For banks classified as heterogeneous, we conducted pairwise post hoc comparisons using Fisher's exact test to examine whether one or more variants differed significantly in difficulty from the remaining variants.  

\subsection{LM-Based Pre-Deployment Evaluation of Isomorphic Problems}

To examine problem quality in the absence of student response data, we used a suite of open-source language models (LMs) to solve all isomorphic problems in 20 problem banks. The current study was limited to banks with numerical answers or multiple-choice formats, because smaller models showed insufficient capability on multiple-answer and categorization questions. As shown in Table~\ref{tab:results}, 13 of the 20 problem banks evaluated by LMs overlapped with banks that were also deployed on exams. Although some of the evaluated banks included image-based questions, none of the LMs used in this study had multi-modal input capability. For those banks, LM performance should therefore be interpreted as reflecting model responses to the text portion of the problem only. Because the LM-based evaluation was conducted retrospectively, student data were collected before any items could be excluded or revised on the basis of LM results.

\subsubsection{Model Selection.}

We evaluated 17 open-source LMs spanning four model families, with model scales from 0.6B to 32B parameters. The model set included base, instruction-tuned, and reasoning-enabled variants. All models were evaluated using hyperparameters specified in their respective model cards.\footnote{Specific hyperparameters can be found at \url{https://huggingface.co/models}.}

\begin{itemize}
\item \textbf{Qwen3}~\cite{yang2025qwen3} (\textbf{9 models}): Qwen3-0.6B, Qwen3-1.7B, Qwen3-4B (Base, Instruct, Thinking), Qwen3-14B, Qwen3-32B, Qwen3-30B-A3B (Instruct, Thinking).
\item \textbf{Llama3}~\cite{dubey2024llama} (\textbf{3 models}): Llama-3.2-1B, Llama-3.2-3B, Llama-3.1-8B.
\item \textbf{Phi-4}~\cite{abdin2024phi} (\textbf{4 models}): Phi-4 (Base, Reasoning), Phi-4-mini (Instruct, Reasoning).
\item \textbf{GPT-oss}~\cite{agarwal2025gpt} (\textbf{1 model}): GPT-oss-20B
\end{itemize}

\subsubsection{Prompting Strategy.}

We used zero-shot prompting with JSON-formatted output to support automated parsing and evaluation. The full prompt is shown below:

 \begin{quote}
 \textit{You are an expert Physics Solver. Your goal is to provide a correct, clear, step-by-step solution to the problem.}

 \textit{Guidelines:}
 \begin{itemize}
 \item \textit{Identify the physical principles involved (e.g., Newton's Laws).}
 \item \textit{Show your algebraic work before plugging in numbers.}
 \item \textit{State the final answer clearly.}
 \end{itemize}

 \textit{Respond with valid JSON format \newline \{"reasoning": "<detailed solution>",  "answer": "<number / choice >"\}}
 \end{quote}

\subsubsection{LM-Based Accuracy.}

Each isomorphic variant was answered once by each of the 17 evaluated LMs. For each question, LM accuracy was defined as the proportion of models that answered the question correctly. Bank-level LM accuracy was then calculated as the average of these variant-level accuracies across all variants in the bank. Lower bank-level LM accuracy indicates that the bank was more difficult for the evaluated model set.

\subsection{Correlation between LM and Student Performance} 

To assess whether LM performance aligned with student performance, we computed both Pearson correlations~\cite{sedgwick2012pearson} and Spearman rank correlations~\cite{zar1972significance} between LM accuracy and student accuracy across isomorphic variants within each bank. Pearson correlation measured the linear association between LM and student accuracies, whereas Spearman correlation measured whether the two sources produced similar rank-orderings of question difficulty. These correlations were calculated only for problem banks that had both LM evaluation and student response data.

\section{Results}
\begin{table}[tp!]
\centering
\small
\setlength{\tabcolsep}{3pt}
\setlength{\abovecaptionskip}{6pt}
\resizebox{\columnwidth}{!}{
\begin{tabular}{c c c c c c c c c c c}
\toprule
\multirow{2}{*}{\textbf{Bank}} 
& \multirow{2}{*}{\textbf{\#Items}} 
& \multicolumn{3}{c}{\textbf{LMs}} 
& \multicolumn{4}{c}{\textbf{Students}} 
& \multirow{2}{*}{\textbf{Pearson}} 
& \multirow{2}{*}{\textbf{Spearman}} \\
\cmidrule(lr){3-5} \cmidrule(lr){6-9}
& 
& \textbf{acc} & \textbf{std} & \textbf{homo}
& \textbf{\#stu} & \textbf{acc} & \textbf{std} & \textbf{homo}
& & \\
\midrule

\rowcolor{gray!20}
\multicolumn{11}{l}{\textit{Numerical Questions}} \\
\midrule
\textit{0-1}  & 10 & 0.835 & 0.281 & \ensuremath{\times} & - & - & - & - & - & - \\
\textit{1-1}\textsuperscript{\dag}  & 10 & 0.829 & 0.098 & \checkmark & - & - & - & - & - & - \\
\textit{1-2}  & 21 & 0.434 & 0.325 & \ensuremath{\times} & - & - & - & - & - & - \\
\textit{2-2}  & 20 & 0.668 & 0.082 & \checkmark & 247 & 0.627 & 0.218 & \checkmark & 0.406 & 0.464 \\
\textit{3-3}  & 20 & 0.524 & 0.108 & \checkmark & 247 & 0.478 & 0.135 & \checkmark & 0.263 & 0.243 \\
\textit{3-4}\textsuperscript{\dag}  & 48 & 0.438 & 0.135 & \checkmark & 242 & 0.690 & 0.262 & \checkmark & 0.147 & 0.119 \\
\textit{4-1}  & 20 & 0.515 & 0.071 & \checkmark & 240 & 0.796 & 0.118 & \checkmark & -0.141 & -0.134 \\
\textit{6-1}  & 20 & 0.415 & 0.251 & \ensuremath{\times} & 240 & 0.483 & 0.207 & \ensuremath{\times} & 0.537 & 0.433 \\
\textit{7-1}  & 35 & 0.766 & 0.099 & \checkmark & 240 & 0.842 & 0.174 & \ensuremath{\times} & 0.357 & 0.315 \\
\textit{8-1}  & 24 & 0.520 & 0.214 & \ensuremath{\times} & 225 & 0.831 & 0.125 & \checkmark & -0.265 & -0.265 \\
\textit{9-2}  & 30 & 0.733 & 0.087 & \checkmark & - & - & - & - & - & - \\
\textit{9-3}  & 30 & 0.641 & 0.070 & \checkmark & - & - & - & - & - & - \\
\textit{9-4}\textsuperscript{\dag}  & 30 & 0.396 & 0.107 & \checkmark & 225 & 0.671 & 0.202 & \checkmark & -0.136 & -0.165 \\
\textit{10-1}  & 12 & 0.569 & 0.038 & \checkmark & 225 & 0.653 & 0.075 & \checkmark & 0.356 & 0.353 \\

\midrule
\rowcolor{gray!20}
\multicolumn{11}{l}{\textit{Multiple Choice Questions}} \\
\midrule
\textit{0-2}\textsuperscript{\dag}  & 10 & 0.506 & 0.097 & \checkmark & 247 & 0.793 & 0.076 & \checkmark & 0.084 & -0.079 \\
\textit{2-1}\textsuperscript{\dag}  & 26 & 0.272 & 0.192 & \ensuremath{\times} & 247 & 0.810 & 0.138 & \checkmark & -0.372 & -0.375 \\
\textit{3-5}  & 15 & 0.651 & 0.075 & \checkmark & 247 & 0.781 & 0.076 & \checkmark & -0.198 & -0.165 \\
\textit{4-2}  & 20 & 0.662 & 0.093 & \checkmark & - & - & - & - & - & - \\
\textit{9-1}\textsuperscript{\dag}  & 26 & 0.518 & 0.228 & \ensuremath{\times} & 225 & 0.742 & 0.189 & \ensuremath{\times} & 0.445 & 0.456 \\
\textit{11-1}  & 20 & 0.765 & 0.125 & \checkmark & - & - & - & - & - & - \\

\midrule
\rowcolor{gray!20}
\multicolumn{11}{l}{\textit{Other Formats}} \\
\midrule
\textit{3-1}  & 23 & - & - & - & 247 & 0.696 & 0.128 & \checkmark & - & - \\
\textit{3-2}  & 23 & - & - & - & 240 & 0.838 & 0.106 & \checkmark & - & - \\
\textit{3-6}  & 24 & - & - & - & 247 & 0.696 & 0.122 & \checkmark & - & - \\
\textit{5-1}  & 25 & - & - & - & 240 & 0.946 & 0.078 & \checkmark & - & - \\
\textit{6-2}  & 28 & - & - & - & 240 & 0.542 & 0.224 & \ensuremath{\times} & - & - \\
\textit{8-2}  & 23 & - & - & - & 225 & 0.667 & 0.196 & \ensuremath{\times} & - & - \\
\textit{10-2}\textsuperscript{\dag}  & 29 & - & - & - & 225 & 0.733 & 0.194 & \checkmark & - & - \\

\bottomrule
\end{tabular}
}
\caption{Bank-level difficulty statistics for isomorphic problem banks. \textit{acc} = mean accuracy; \textit{std} = standard deviation of item-level accuracies; \textit{\#stu} = number of students responding per bank. LM accuracy is averaged across all 17 models. \textit{Homo} indicates the homogeneity classification from the extended Fisher's exact test at $\alpha = 0.05$ (\checkmark = homogeneous, \ensuremath{\times} = heterogeneous). \textit{Pearson} and \textit{Spearman} denote the Pearson and Spearman rank correlations between item-level LM and student accuracies, respectively. Superscript $\dag$ marks banks that contain image-based questions.}
\label{tab:results}
\end{table}

\subsection{Post-deployment Evaluation using Student Data}

\subsubsection{Bank-level Difficulty.}

Table~\ref{tab:results} summarizes student performance for the 20 problem banks deployed across the three midterm exams. Bank-level correct rates range from 47.8\% to 94.6\%, with an average correct rate of 71.4\%. The lowest-performing banks are \textit{3-3} and \textit{6-1}, with accuracy of 47.8\% and 48.3\%, respectively, whereas the highest-performing bank was \textit{5-1}, with a correct rate of 94.6\%. These results indicate that the deployed banks spanned a broad but appropriate range of difficulty, with most banks remaining solvable while avoiding uniformly near-ceiling performance.

\subsubsection{Homogeneity of Difficulty within Problem Banks.}

Across the 20 deployed banks, the standard deviation of item-level accuracy ranged from 0.076 to 0.262, with an average of 0.154. Eleven of the 20 banks had standard deviations below 0.15, suggesting that more than half of the deployed banks showed relatively small variation in student performance across isomorphic variants. The extended Fisher's exact test at the $\alpha = 0.05$ level classified 15 of the 20 banks (75\%) as homogeneous and 5 as heterogeneous. This result suggests that most generated banks maintained comparable difficulty across variants, although some banks still showed statistically detectable heterogeneity. For the heterogeneous banks, post-hoc comparisons with adjusted p-values did not identify specific variants that differed significantly from the rest of the bank. This result should be interpreted cautiously, since the statistical power of the post-hoc tests was limited and each isomorphic variant received only between 5 and 20 student responses.

\subsection{Pre-deployment evaluation using LM simulation}

\subsubsection{Bank-level LM Difficulty.}

Table~\ref{tab:results} summarizes LM performance across the 20 problem banks evaluated through simulation. Bank-level LM accuracy ranged from 0.272 to 0.835, with an average accuracy of 0.583 across banks. This wide range suggests that LM-based simulation can distinguish relatively easy and difficult banks before classroom deployment.

\subsubsection{LM-based Detection of Heterogeneous Banks.}

The standard deviation of LM performance within each bank ranged from 0.038 to 0.325, with an average of 0.139. Ten of the 20 banks had LM standard deviations below 0.10, suggesting relatively consistent LM performance across variants. In contrast, five banks had standard deviations above 0.20, which could warrant instructor review before classroom deployment. The extended Fisher's exact test classified 14 of the 20 banks (70\%) as homogeneous and 6 as heterogeneous. Most banks with high LM variance were also classified as heterogeneous by Fisher's exact test, suggesting that LM simulation can provide a useful pre-deployment signal for identifying banks that may contain unintended difficulty variation.

\subsection{Correlation between LM Simulation and Student Performance}

\subsubsection{Question-Level Correlation Analysis.}

Table~\ref{tab:results} reports the question-level correlations between LM and student accuracy for the 13 banks with both LM and student data. Pearson correlations range from $-0.372$ to $0.537$, while Spearman correlations range from $-0.375$ to $0.464$. Five of the 13 banks showed moderate positive alignment between LM and student performance, with Pearson correlations above $0.30$. These banks included both numerical and multiple-choice problem banks, suggesting that LM simulation can capture student difficulty patterns for some, but not all, isomorphic banks.

The alignment is stronger for banks with moderate difficulty and sufficient variation in student performance across variants. In contrast, weak or negative correlations appear in banks where student performance was near ceiling, where LM and student difficulty patterns diverge, or where image information may carry essential content. In particular, image-based banks show weaker LM-student alignment on average: among the five image-based banks with correlation data, the mean Pearson and Spearman correlations were $0.034$ and $-0.009$, respectively, compared with $0.164$ and $0.156$ for the remaining banks.

\subsubsection{Alignment in Identifying Heterogeneous Problem Banks.}

LM simulation identified 6 of the 20 evaluated problem banks as having heterogeneous difficulty across variants. Of the 13 banks with both LM and student data, student response data flagged 3 banks (\textit{6-1}, \textit{7-1}, \textit{9-1}) as heterogeneous. LM simulation correctly flagged 2 of these (\textit{6-1}, \textit{9-1}) but missed \textit{7-1}. LM simulation also flagged \textit{2-1} and \textit{8-1} as heterogeneous, although the student data did not confirm this. Overall, these results suggest that LM simulation can serve as a useful but imperfect screening tool for identifying banks that may require closer review before deployment. Its value lies primarily in flagging potential heterogeneity rather than definitively classifying banks as problematic.

\subsection{Case Study: Diagnosing Outlier Problems}
To investigate whether LM validation can identify specific problematic versions within a bank, we conducted case studies on Bank \textbf{6-1} and \textbf{9-1}, both of which exhibit heterogeneous difficulty in both student and LM performance (Table~\ref{tab:results}). 

\subsubsection{Case Study 1: Bank \textbf{6-1}.}

For bank \textbf{6-1}, LM evaluation flagged six items as having abnormally low accuracy relative to the bank average: \textit{q2, q5, q16, q17, q19, q20}, as shown in Figure~\ref{fig:case_study}. Three of these items (\textit{q16, q17, q20}) also had the lowest accuracy in the student data, suggesting that LM validation successfully identified genuine difficulty outliers. Problem \textit{q19} had too few student responses for a reliable difficulty estimate, whereas the remaining two flagged problems (\textit{q2, q5}) did not show significantly different student performance from the rest of the bank. To better understand why these items were flagged, we qualitatively examined the responses generated by Qwen3-32B. This analysis revealed three distinct categories of LM failure:

\begin{figure}[t!]
    \includegraphics[width = 1\textwidth]{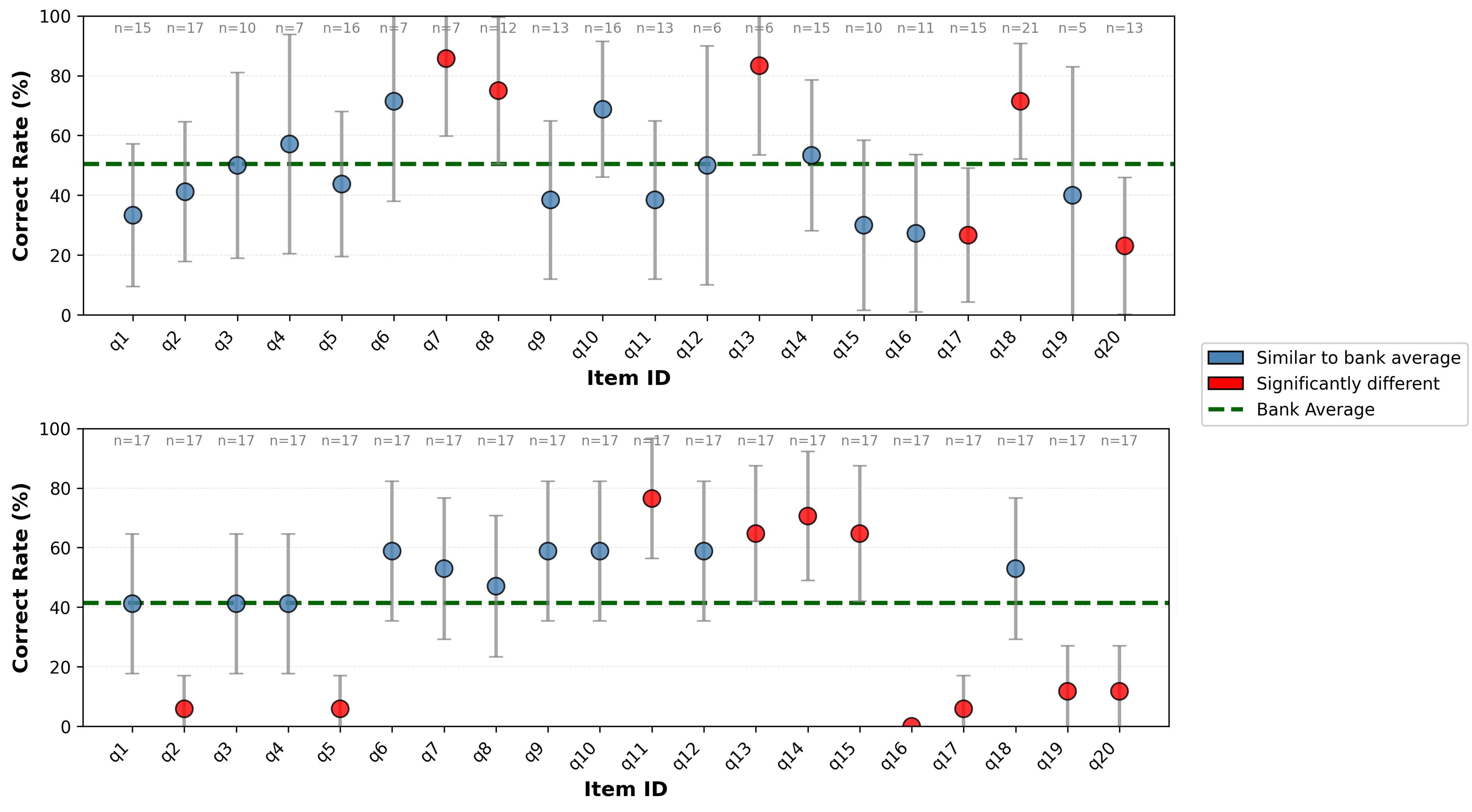}

    \caption{Comparison of student accuracy distribution (Top) and LM simulated accuracy (Bottom) of problem bank 6-1. Error bars represent standard error of measurement.}
    \label{fig:case_study}
\end{figure}

\subsubsection*{Category 1: Insufficient World Knowledge (\textit{q16, q17, q19})}
In this category, the LM failed to apply implicit real-world knowledge needed to correctly model the physical situation. The issue is illustrated by \textit{q17}:

\par\medskip
\fbox{
    \parbox{0.9\linewidth}{ 
        \small
        A small boat is coasting toward a dock. To stop it gently, an elastic bungee net is stretched across the path. The boat travels a short distance through the water, experiencing drag, before contacting the net and stretching it to a stop. Given: The boat's mass is 250 kg and its initial velocity is 3.0 m/s. It coasts for 5.0 m before hitting the net. The constant drag force from the water is 150 N. The net stretches by 2.0 m to bring the boat to rest. What is the effective spring constant ($k$) of the bungee net in N/m?
    }
}
\par\medskip

The LM failed to incorporate the fact that the boat is still traveling in the water when it touches the net. Therefore, drag continues to act while the net stretches, and the additional $2.0$ m stretching distance should be included when calculating the work done by the water resistance.

\subsubsection*{Category 2: Overlooked Verbal Condition (\textit{q20})}

In this item, the LM disregarded a key verbal cue embedded in the problem statement:

\par\medskip
\fbox{
    \parbox{0.9\linewidth}{ 
        \small
        A ball is rolling with some initial speed. It enters a $2.0$ m long patch of mud, which provides a constant resistive force. Nearing the middle of the mud patch, it hits an elastic band that stretches until the ball stops.
    }
}
\par\medskip

The LM failed to recognize that the phrase ``Nearing the middle of the mud patch'' means that the ball only traveled about $1.0$ m before hitting the elastic band, and instead mistakenly used the full $2.0$ m length of the mud patch. It also failed to understand that the elastic band was located within the mud patch, due to lack of world knowledge.

\subsubsection*{Category 3: Ambiguous Problem Text (\textit{q2, q5})}

In these items, the problem statement omitted or under-specified a key condition. For example, \textit{q5} states:

\par\medskip
\fbox{
    \parbox{0.9\linewidth}{ 
        \small
        For a movie scene, a stunt performer executes a high fall from a platform onto a large, soft crash mat below. As the performer lands, the mat compresses, exerting a constant upward force to safely slow them down. Given: The performer falls from a height of $15$ m. The mat compresses by a distance of $1.5$ m while providing an upward force of $5000$ N. 
    }
}
\par\medskip

The problem body failed to specify whether the $15$ m height was measured from the ground or from the surface of the compressed mat. This ambiguity led the LM to interpret the problem in a way that differed from the intended solution. Notably, the course instructor reported that students raised the same question during the exam, leading to a class-wide clarification.

Categories 1 and 2 correspond to genuine difficulty outliers that are also reflected in the student data. The LM failures in these categories provide plausible explanations for why students may have found these items difficult, suggesting that the errors reflect shared sources of confusion rather than model-specific artifacts. In contrast, Category 3 highlights ambiguities in the problem text itself, which are flaws that may not appear in student performance data if they were clarified during exam administration.

\subsubsection{Case Study 2: Bank \textbf{9-1}.}

In contrast to bank \textbf{6-1}, bank \textbf{9-1} exhibited a cleaner and more consistent pattern. This problem bank consists of multiple-choice problems that ask students to calculate the moment of inertia of composite objects made up of a rod, a second object attached (ring, disk or square plate), and a point mass. Three problems (\textit{q20, q23, q26}) that had exceptionally low student accuracy (20\% lower than other problems) were also among the four problems that had the lowest LM accuracy. The problem body of \textit{q23} is shown below as a representative example:

\par\medskip
\fbox{
    \parbox{0.9\linewidth}{ 
        \small
        A composite object consists of a $\textbf{uniform thin rod}$ of mass $m_1$ and length $L$, pivoted about its $\textbf{right end}$ (indicated by the black dot in the figure, at position $L$). A $\textbf{thin ring}$ of mass $m_2$ and diameter $d$ is rigidly attached to this pivot point so that the rod passes through the center of the ring. A $\textbf{small point mass}$ of mass $m_3$ (shown as the $\textbf{green circular dot}$) is fixed to the $\textbf{midpoint}$ of the rod (at $L/2$). The composite object rotates in a vertical plane about the pivot point at the right end of the rod. Which of the following expressions correctly represents the $\textbf{moment of inertia $I$}$ of this composite object about the pivot?

        \begin{center}
            \includegraphics[width=0.35\linewidth]{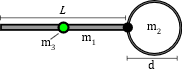}
        \end{center}
    }
}
\par\medskip

All three problems share a nearly identical geometric configuration: the second object is a thin ring, and the pivot is located between the rod and the ring. Across the LMs examined, the most common errors were either selecting an option that ignored the parallel-axis theorem or applying it incorrectly. This pattern suggests that this specific configuration, a thin ring with the pivot located on its edge, makes students more likely to overlook the need for the parallel-axis theorem.

Overall, the two case studies support the use of LM simulation as a diagnostic tool for identifying both difficulty heterogeneity and text-level deficiencies in isomorphic problem banks prior to deployment.

\section{Discussion}
This study investigates whether GenAI can support the scalable creation of isomorphic problem banks for introductory physics, whether the generated banks exhibit acceptable consistency in difficulty when evaluated with student performance data, and whether LM-based simulation can function as a pre-deployment quality control tool. Overall, the results suggest that the proposed workflow is a practical approach for the scalable isomorphic problem banks construction. Most generated banks show acceptable difficulty homogeneity in student data, while LM-based simulation provides a useful, low-cost method for flagging variants that may require closer human review before classroom deployment.

\myparagraph{RQ1: To what extent are item difficulties homogeneous within banks generated by the current framework?}

Both student data and LM simulation provide evidence that the framework can generate isomorphic problem banks with acceptable consistency in difficulty. In the student-data analysis, 75\% of the banks are classified as homogeneous by Fisher's exact test, with a mean within-bank standard deviation of 0.154 in item accuracy. LM simulation produces a similar pattern: 70\% of the evaluated banks are classified as homogeneous, with an average within-bank standard deviation of 0.139. Together, these findings suggest that the proposed framework generally preserves difficulty consistency across isomorphic variants, although a minority of banks still show statistically detectable heterogeneity and require closer human review.

In addition, across the 20 deployed banks, student accuracy ranges from 47.8\% to 94.6\%, with an average of 71.4\%, which, according to the instructor, is comparable to the accuracy of similar problems deployed on traditional midterm exams in the same course in previous semesters. This supports the validity of the new problem-bank-based assessment format, in which students are given access to the full problem banks for practice before the exam. A likely reason for this consistency is that the framework separates structural and contextual variation during problem creation, applies different tools to distinct generation tasks, and incorporates opportunities for human review and revision at each stage of the AI-assisted workflow. This process is also highly reproducible in practice: it can be carried out using commercially available GenAI services, and the prompts used for generation are recorded alongside the problem statements in the dataset.

\myparagraph{\textbf{RQ2: To what extent is LM performance aligned with student performance on the same banks?}}

We observe varying levels of alignment between LM and student performance at the individual-problem level. Alignment is strongest for banks with moderate difficulty, multi-step numerical reasoning, no image-based information, and sufficient within-bank variation for relative difficulty differences to be detectable. Under these conditions, LM simulation appears to be useful in estimating the relative difficulty of problems within a bank.

The results also show that LM-based validation is sensitive to model choice. Very small models exhibit floor effects, whereas larger models approach ceiling performance, limiting their usefulness for distinguishing among isomorphic variants. Mid-to-large instruction-tuned or reasoning-based models appear to be the most useful because their performance overlaps more closely with the student accuracy observed in this study. This finding has an important practical implication: effective validation does not require the strongest available model, but rather a model whose level of competence produces meaningful variation across problem variants.

LM simulation is also useful for identifying banks with significantly heterogeneous difficulty. Among the banks with both LM and student data, LM simulation successfully flags two of the three banks that are found to be heterogeneous based on student performance, while also flagging two banks whose heterogeneity is not confirmed by student data. Several factors may explain this mismatch. First, the student-side statistical tests have limited power because each isomorphic item receives only a small number of responses, so some genuinely heterogeneous banks may have gone undetected. Second, because the banks are available to students for practice before the exam, students may become familiar with certain contextual variations, thereby reducing performance differences that might otherwise appear. Third, in Bank \textit{2-1}, the figure contains information essential for solving the problem, placing text-only LM simulation at an inherent disadvantage. Overall, these results suggest that LM simulation can serve as a useful early-warning tool for identifying banks that warrant closer human review.

Beyond identifying heterogeneity within problem banks, LM simulation also provides a rapid estimate of relative bank-level difficulty. LMs and students often agree on which banks are comparatively challenging or accessible, especially for numerical questions. Estimating bank-level difficulty can be useful for instructional planning, for example, by helping instructors select easier banks for open practice and reserve more demanding banks for formal evaluation. In conventional settings, developing comparable difficulty profiles through pilot testing alone would require repeated deployments across multiple terms.

\myparagraph{\textbf{RQ3: Can LM responses help identify the likely causes of heterogeneous difficulty?}}

The case studies suggest that LM responses can provide useful information for improving the homogeneity of problem banks. For both heterogeneous banks examined in detail, the problems identified as difficult in LM simulation overlapped substantially with the problems that are more difficult for students. From an instructional perspective, the LM errors also resemble plausible student errors rather than artifacts of weak model performance. In Bank \textit{6-1}, LM responses reveal failures to apply relevant implicit knowledge, failures to notice key verbal conditions, and confusion caused by under-specified wording. In Bank \textit{9-1}, the recurring error involves omitting the parallel-axis theorem for a particular geometric configuration. These patterns are meaningful diagnostic signals because they point to concrete reasons why certain variants may be harder than intended.

LM analysis is also valuable because it surfaces revision targets that are not fully visible from student performance alone. In Bank \textit{6-1}, for example, the simulation exposes under-specified wording in two problems that do not clearly stand out in the student statistics. Taken together, the case studies show that LM-based analysis can contribute at two levels: it can identify which banks or items deserve closer inspection, and generate actionable recommendations on how to revise them.

Overall, the findings suggest that the current prompt-chain framework can produce isomorphic problem banks at scale, with substantial contextual variation while maintaining an adequate level of homogeneity in difficulty. These banks, in turn, enable an assessment format in which problem banks are made available to students in advance for practice before being used in exams. At the same time, LM simulation shows promise as a practical method for identifying potentially problematic isomorphic variants and suggesting directions for revision before deployment. Taken together, these two components point toward a feasible workflow for AI-assisted assessment development, in which scalable generation is paired with low-cost pre-deployment quality control.

\section{Limitations and Future Work}
In this section, we discuss several key limitations of the current study and outline possible directions for addressing them in future work.

First, with respect to the problem-generation framework, the use of a single prompt chain is convenient and broadly accessible to instructors using commercially available GenAI services. However, when the model is asked to generate large numbers of isomorphic variants or generate a large bank in a single pass, the quality and consistency of the outputs tend to decline. This likely reflects a combination of practical constraints, including prompt complexity, context management, and output-length limitations. In addition, some steps in the prompt chain, such as data-format conversion, are repetitive and could be automated further. Consequently, these limitations motivate the development of an agentic problem-generation framework that can both design and execute the prompt chain for an entire problem bank under human supervision. Such a framework could also generate and validate problem variants in smaller, coordinated batches, reducing the quality loss that arises in large one-shot generations. In this workflow, the instructor would only need to provide the original problem and, optionally, a small number of example isomorphic variations, while moderating the output at each step.

Second, the current problem generation framework is limited to generating relatively simple diagrams that can be specified directly from mathematical formulas, such as parabolic trajectories. An important direction for future work is to extend the workflow to support the reliable and accurate generation of more complex images and diagrams.

Third, in evaluating problem quality, we used problem difficulty as the sole proxy for quality and did not examine other important factors, such as the readability of the problem text or potential bias in contextual framing and distractor design. Although some of these issues may be partially reflected in difficulty data, they are likely to remain undetected given the small sample size and limited statistical power of the present study. In addition, the LM simulations are unlikely to detect such issues, especially when performance is evaluated only in terms of correctness. Therefore, a natural next step is to develop dedicated models or AI agents that can provide critical insight into these dimensions. Such agents could then be integrated into the workflow in a manner similar to the multi-agent framework proposed in \cite{researchgate2025lmaig}.

Fourth, the current LM-based pre-deployment evaluation framework remains limited in both model selection and calibration. In this study, we used a cohort of general-purpose LMs with different parameter sizes to emulate students of different ability levels. However, our results suggest that smaller models tended to exhibit floor effects, whereas larger models tended to exhibit ceiling effects in terms of correctness, reducing their usefulness for evaluation. This finding suggests that future work should examine whether comparable results can be achieved with a smaller set of carefully selected medium-sized models. In addition, all models in the current study were evaluated using a single generic prompt. Although prior research has shown similarities between the performance of small LMs and introductory-level students~\cite{liu2025llms,lu2024generative}, future work could examine whether one or more models can be specifically prompted or trained to better match the response patterns of students at particular ability levels, or of specific student populations, thereby improving the validity of pre-deployment evaluation.

Fifth, the current evaluation pipeline remains limited in its ability to handle certain problem formats. Problems containing images are especially difficult for the text-based LMs used in this study, and multiple-choice banks also appear to be less amenable to LM-based validation than numerical problems. Future work should therefore extend the method to multimodal problems and explore more robust approaches for evaluating a wider range of answer formats, including multiple-answer and categorization questions.

Finally, in the current study, LM-based evaluation is conducted as a separate step after problem creation. In future work, however, LM evaluation could be integrated directly into the problem-generation workflow to support iterative generation of problem variants and improve generation quality. In the longer term, such a combined workflow could make real-time creation of isomorphic problems more feasible. For example, one could imagine a system that uses the problems a student has previously practiced to generate new isomorphic assessment problems of comparable difficulty at the start of an exam. Such a system could enable truly personalized assessment at scale.

\begin{credits}


\end{credits}
%
%
%
\bibliographystyle{splncs04}
\bibliography{references}

@article{kurdi2020systematic,
  title={A systematic review of automatic question generation for educational purposes},
  author={Kurdi, Ghader and Leo, Jared and Parsia, Bijan and Sattler, Uli and Al-Emari, Salam},
  journal={International journal of artificial intelligence in education},
  volume={30},
  number={1},
  pages={121--204},
  year={2020},
  publisher={Springer}
}

@article{kosh2019cost,
  title={A cost--benefit analysis of automatic item generation},
  author={Kosh, Audra E and Simpson, Mary Ann and Bickel, Lisa and Kellogg, Mark and Sanford-Moore, Ellie},
  journal={Educational Measurement: Issues and Practice},
  volume={38},
  number={1},
  pages={48--53},
  year={2019},
  publisher={Wiley Online Library}
}

@article{chan2025automatic,
  title={Automatic item generation in various STEM subjects using large language model prompting},
  author={Chan, Kuang Wen and Ali, Farhan and Park, Joonhyeong and Sham, Kah Shen Brandon and Tan, Erdalyn Yeh Thong and Chong, Francis Woon Chien and Qian, Kun and Sze, Guan Kheng},
  journal={Computers and Education: Artificial Intelligence},
  volume={8},
  pages={100344},
  year={2025},
  publisher={Elsevier}
}

@article{omopekunola2024automatic,
  title={Automatic generation of physics items with large language models (LLMs)},
  author={Omopekunola, Moses Oluoke and Kardanova, Elena Yu},
  journal={REID (Research and Evaluation in Education)},
  volume={10},
  number={2},
  pages={4},
  year={2024}
}

@article{maity2025can,
  title={Can large language models meet the challenge of generating school-level questions?},
  author={Maity, Subhankar and Deroy, Aniket and Sarkar, Sudeshna},
  journal={Computers and Education: Artificial Intelligence},
  volume={8},
  pages={100370},
  year={2025},
  publisher={Elsevier}
}

@article{arendasy2007using,
  title={Using psychometric technology in educational assessment: The case of a schema-based isomorphic approach to the automatic generation of quantitative reasoning items},
  author={Arendasy, Martin and Sommer, Markus},
  journal={Learning and Individual Differences},
  volume={17},
  number={4},
  pages={366--383},
  year={2007},
  publisher={Elsevier}
}

@article{millar2021repeat,
  title={Repeat individualized assessment using isomorphic questions: a novel approach to increase peer discussion and learning},
  author={Millar, Russell and Manoharan, Sathiamoorthy},
  journal={International Journal of Educational Technology in Higher Education},
  volume={18},
  number={1},
  pages={22},
  year={2021},
  publisher={Springer}
}

@article{norberg2023rewriting,
  title={Rewriting Math Word Problems with Large Language Models.},
  author={Norberg, Kole and Almoubayyed, Husni and Fancsali, Stephen E and De Ley, Logan and Weldon, Kyle and Murphy, April and Ritter, Steve},
  journal={Grantee Submission},
  year={2023},
  publisher={ERIC}
}

@article{yang2025qwen3,
  title={Qwen3 technical report},
  author={Yang, An and Li, Anfeng and Yang, Baosong and Zhang, Beichen and Hui, Binyuan and Zheng, Bo and Yu, Bowen and Gao, Chang and Huang, Chengen and Lv, Chenxu and others},
  journal={arXiv preprint arXiv:2505.09388},
  year={2025}
}

@article{dubey2024llama,
  title={The llama 3 herd of models},
  author={Dubey, Abhimanyu and Jauhri, Abhinav and Pandey, Abhinav and Kadian, Abhishek and Al-Dahle, Ahmad and Letman, Aiesha and Mathur, Akhil and Schelten, Alan and Yang, Amy and Fan, Angela and others},
  journal={arXiv e-prints},
  pages={arXiv--2407},
  year={2024}
}

@article{abdin2024phi,
  title={Phi-4 technical report},
  author={Abdin, Marah and Aneja, Jyoti and Behl, Harkirat and Bubeck, S{\'e}bastien and Eldan, Ronen and Gunasekar, Suriya and Harrison, Michael and Hewett, Russell J and Javaheripi, Mojan and Kauffmann, Piero and others},
  journal={arXiv preprint arXiv:2412.08905},
  year={2024}
}

@article{agarwal2025gpt,
  title={gpt-oss-120b \& gpt-oss-20b model card},
  author={Agarwal, Sandhini and Ahmad, Lama and Ai, Jason and Altman, Sam and Applebaum, Andy and Arbus, Edwin and Arora, Rahul K and Bai, Yu and Baker, Bowen and Bao, Haiming and others},
  journal={arXiv preprint arXiv:2508.10925},
  year={2025}
}

@article{muldoon2012,
  author = {Muldoon, R.},
  title = {Is it time to ditch the traditional university exam?},
  journal = {Higher Education Research and Development},
  volume = {31},
  number = {2},
  pages = {263--265},
  year = {2012},
  doi = {10.1080/07294360.2012.680249}
}

@inproceedings{zilles2019,
  author = {Zilles, C. and West, M. and Mussulman, D. and Bretl, T.},
  title = {Making testing less trying: Lessons learned from operating a computer-based testing facility},
  booktitle = {Proceedings - Frontiers in Education Conference (FIE)},
  year = {2019},
  doi = {10.1109/FIE.2018.8658551}
}

@article{Flugge2024,
  author  = {Flugge, Valerie},
  title   = {Cheaters Never Prosper: The Legal Liability and Ethical Responsibility of ``Homework Help'' Sites},
  journal = {Notre Dame Journal of Law, Ethics \& Public Policy},
  year    = {2024},
  volume  = {38},
  number  = {1},
  pages   = {177--212}
}

@article{Francis2025,
  author  = {Francis, Nigel J. and Jones, Sue and Smith, David P.},
  title   = {Generative {AI} in Higher Education: Balancing Innovation and Integrity},
  journal = {British Journal of Biomedical Science},
  year    = {2025},
  volume  = {81},
  doi     = {10.3389/bjbs.2024.14048}
}

@article{Newton2025,
  author  = {Newton, Philip M. and Draper, Michael J.},
  title   = {Widespread use of summative online unsupervised remote ({SOUR}) examinations in {UK} higher education: ethical and quality assurance implications},
  journal = {Quality in Higher Education},
  year    = {2025},
  volume  = {31},
  number  = {1},
  pages   = {127--141},
  doi     = {10.1080/13538322.2025.2521174}
}

@inproceedings{Frederick2024PERC,
  author    = {Frederick, Emily and Chen, Zhongzhou},
  title     = {Comparing student performance on a multi-attempt asynchronous assessment to a single-attempt synchronous assessment in introductory level physics},
  booktitle = {2024 Physics Education Research Conference Proceedings},
  year      = {2024},
  pages     = {148--153},
  doi       = {10.1119/perc.2024.pr.Frederick},
  note      = {Also available as arXiv:2407.15257}
}

@inproceedings{Silva2020SIGCSE,
  author    = {Silva, Mariana and West, Matthew and Zilles, Craig},
  title     = {Measuring the Score Advantage on Asynchronous Exams in an Undergraduate CS Course},
  booktitle = {Proceedings of the 51st ACM Technical Symposium on Computer Science Education},
  year      = {2020},
  pages     = {873--879},
  doi       = {10.1145/3328778.3366859}
}

@inproceedings{Sud2019ASEE,
  author    = {Sud, Paras and West, Matthew and Zilles, Craig},
  title     = {Reducing Difficulty Variance in Randomized Assessments},
  booktitle = {2019 ASEE Annual Conference \& Exposition},
  year      = {2019},
  url       = {https://peer.asee.org/33234}
}

@article{Sullivan2016,
  author    = {Sullivan, Daniel P.},
  title     = {An Integrated Approach to Preempt Cheating on Asynchronous, Objective, Online Assessments in Graduate Business Classes},
  journal   = {Online Learning},
  year      = {2016},
  volume    = {20},
  number    = {3},
  pages     = {195--209},
  doi       = {10.24059/olj.v20i3.650}
}

@article{Liu2025Frontiers,
  author  = {Liu, Chang and Xie, Rui and Chen, Zhongzhou},
  title   = {Towards actionable recommendations for exam preparation using isomorphic problem banks and Explainable Machine Learning},
  journal = {Frontiers in Education},
  year    = {2025},
  volume  = {10},
  pages   = {1632132},
  doi     = {10.3389/feduc.2025.1632132}
}

@InProceedings{Jiao2023AutomaticEQ,
author="Jiao, Ying
and Shridhar, Kumar
and Cui, Peng
and Zhou, Wangchunshu
and Sachan, Mrinmaya",
editor="Wang, Ning
and Rebolledo-Mendez, Genaro
and Matsuda, Noboru
and Santos, Olga C.
and Dimitrova, Vania",
title="Automatic Educational Question Generation with Difficulty Level Controls",
booktitle="Artificial Intelligence in Education",
year="2023",
publisher="Springer Nature Switzerland",
address="Cham",
pages="476--488"
}

@inproceedings{Dijkstra2022ReadingCQ,
  title={Reading Comprehension Quiz Generation using Generative Pre-trained Transformers},
  author={Ramon Dijkstra and Z{\"u}lk{\"u}f Genç and Subhradeep Kayal and J. Kamps},
  booktitle={iTextbooks@AIED},
  year={2022},
  url={https://api.semanticscholar.org/CorpusID:251771032}
}

@inproceedings{lu2024generative,
  title={Generative students: Using llm-simulated student profiles to support question item evaluation},
  author={Lu, Xinyi and Wang, Xu},
  booktitle={Proceedings of the Eleventh ACM Conference on Learning@ Scale},
  pages={16--27},
  year={2024}
}

@inproceedings{scarlatos2025smart,
  title={Smart: Simulated students aligned with item response theory for question difficulty prediction},
  author={Scarlatos, Alexander and Fernandez, Nigel and Ormerod, Christopher and Lottridge, Susan and Lan, Andrew},
  booktitle={Proceedings of the 2025 Conference on Empirical Methods in Natural Language Processing},
  pages={25082--25105},
  year={2025}
}

@inproceedings{liu2025llms,
  title={Do LLMs Make Mistakes Like Students? Exploring Natural Alignments Between Language Models and Human Error Patterns},
  author={Liu, Naiming and Sonkar, Shashank and Baraniuk, Richard},
  booktitle={International Conference on Artificial Intelligence in Education},
  pages={364--377},
  year={2025},
  organization={Springer}
}

@article{upton1992fisher,
  title={Fisher's exact test},
  author={Upton, Graham JG},
  journal={Journal of the Royal Statistical Society: Series A (Statistics in Society)},
  volume={155},
  number={3},
  pages={395--402},
  year={1992},
  publisher={Wiley Online Library}
}

@article{sedgwick2012pearson,
  title={Pearson’s correlation coefficient},
  author={Sedgwick, Philip},
  journal={Bmj},
  volume={345},
  year={2012},
  publisher={British Medical Journal Publishing Group}
}

@article{achiam2023gpt,
  title={Gpt-4 technical report},
  author={Achiam, Josh and Adler, Steven and Agarwal, Sandhini and Ahmad, Lama and Akkaya, Ilge and Aleman, Florencia Leoni and Almeida, Diogo and Altenschmidt, Janko and Altman, Sam and Anadkat, Shyamal and others},
  journal={arXiv preprint arXiv:2303.08774},
  year={2023}
}

@article{jmir2026anesthesiology,
  title={Fine-Tuned Large Language Models for Generating Multiple-Choice Questions in Anesthesiology: Psychometric Comparison With Faculty-Written Items},
  author={H{\"o}lzing, Carlos Ramon and Meynhardt, Charlotte and Meybohm, Patrick and K{\"o}nig, Sarah and Kranke, Peter},
  journal={JMIR Formative Research},
  volume={10},
  pages={e84904},
  year={2026}
}

@article{deepquestion2025arxiv,
  title={DeepQuestion: Systematic Generation of Real-World Challenges for Evaluating LLMs Performance},
  author={Khoramfar, Ali and Ramezani, Ali and Mohajeri, Mohammad Mahdi and Dousti, Mohammad Javad and Ahmadabadi, Majid Nili and Faili, Heshaam},
  journal={arXiv preprint arXiv:2505.24532},
  year={2025}
}

@article{researchgate2025lmaig,
  title={AI-powered Automatic Item Generation for Psychological Tests: A Conceptual Framework for an LLM-based Multi-Agent AIG System},
  author={Lee, Philseok and Son, Mina and Jia, Zihao},
  journal={Journal of Business and Psychology},
  volume={41},
  pages={71--99},
  year={2025},
  publisher={Springer Nature}
}

@article{zar1972significance,
  title={Significance testing of the Spearman rank correlation coefficient},
  author={Zar, Jerrold H},
  journal={Journal of the American Statistical Association},
  volume={67},
  number={339},
  pages={578--580},
  year={1972},
  publisher={Taylor \& Francis}
}

@article{hambleton1993comparison,
  title={Comparison of classical test theory and item response theory and their applications to test development},
  author={Hambleton, Ronald K and Jones, Russell W},
  journal={Educational measurement: issues and practice},
  volume={12},
  number={3},
  pages={38--47},
  year={1993}
}

@article{peters2025text,
  title={Text-Based Approaches to Item Difficulty Modeling in Large-Scale Assessments: A Systematic Review},
  author={Peters, Sydney and Zhang, Nan and Jiao, Hong and Li, Ming and Zhou, Tianyi and Lissitz, Robert},
  journal={arXiv preprint arXiv:2509.23486},
  year={2025}
}

@inproceedings{yaneva2024findings,
  title={Findings from the first shared task on automated prediction of difficulty and response time for multiple-choice questions},
  author={Yaneva, Victoria and North, Kai and Baldwin, Peter and Ha, Le An and Rezayi, Saed and Zhou, Yiyun and Choudhury, Sagnik Ray and Harik, Polina and Clauser, Brian},
  booktitle={Proceedings of the 19th Workshop on Innovative Use of NLP for Building Educational Applications (BEA 2024)},
  pages={470--482},
  year={2024}
}

@inproceedings{rogoz2024unibucllm,
  title={Unibucllm: Harnessing LLMs for automated prediction of item difficulty and response time for multiple-choice questions},
  author={Rogoz, Ana-Cristina and Ionescu, Radu Tudor},
  booktitle={Proceedings of the 19th Workshop on Innovative Use of NLP for Building Educational Applications (BEA 2024)},
  pages={493--502},
  year={2024}
}

@inproceedings{veeramani2024large,
  title={Large language model-based pipeline for item difficulty and response time estimation for educational assessments},
  author={Veeramani, Hariram and Thapa, Surendrabikram and Shankar, Natarajan Balaji and Alwan, Abeer},
  booktitle={Proceedings of the 19th Workshop on Innovative Use of NLP for Building Educational Applications (BEA 2024)},
  pages={561--566},
  year={2024}
}

@inproceedings{park2024large,
  title={Large language models are students at various levels: Zero-shot question difficulty estimation},
  author={Park, Jae-Woo and Park, Seong-Jin and Won, Hyun-Sik and Kim, Kang-Min},
  booktitle={Findings of the Association for Computational Linguistics: EMNLP 2024},
  pages={8157--8177},
  year={2024}
}

\newpage
\appendix
\section{Problem Bank Generation Prompts}
\label{app:a}

The following prompts were used sequentially to generate the problem bank described in Section~\ref{sec:framework}. The prompts were originally stored and managed in a YAML configuration file and have been reproduced here with minor edits for readability.

\begin{promptbox}{Prompt 1: Initial Scenario Generation}
I'm writing physics problems involving calculation of kinetic friction and balance of forces on a moving object. As a first step, please help me by doing the following:

Generate 10 different cases of an object being pulled or pushed by either a human, an animal, or a machine/vehicle across a surface. Indicate whether the pushing or pulling force must be angled upwards or downwards. 

\medskip
\noindent\textbf{Here are some examples: }
\begin{itemize}
    \item A horse pulling a sledge on snow. Upwards.
    \item A person pushing a couch across carpeted floor. Downwards. 
    \item A person dragging a heavy luggage case. Upwards.
    \item A towing truck dragging a damaged car on rough road. Upwards.
\end{itemize}
\end{promptbox}

\begin{promptbox}{Prompt 2: Variable Generation and Validation}
Now for each of those 10 cases, first generate the following set of values.

\begin{enumerate}
    \item Coefficient of kinetic friction, $\mu_k$. Between 0.1 and 0.9.
    \item Angle of the force $\theta$. Positive 10 degrees to 60 degrees if the force is upwards, negative 10 degrees to 60 degrees if the force is downwards.
    \item Mass of the object being pulled or pushed, M. Appropriate for the object being pushed or pulled, in units of kg.
    \item In addition, generate a generous estimation of the force that can be exerted by the human, animal or machine doing the pushing or pulling (not the object).
\end{enumerate}

Next, calculate the magnitude of the force according to:

\[
F = \frac{\mu * m * g}{\cos(\theta) + \mu * \sin(\theta)}
\]

Check if the magnitude of $F$ (in Newtons) is within the estimated range, and if the magnitude is positive. If yes, list the values. If not, re-generate the random numbers.
\end{promptbox}

\begin{promptbox}{Prompt 3: Problem Statement Drafting}
For each of the 10 scenarios and the corresponding set of numbers, write a physics problem following those steps:

\begin{enumerate}
    \item Choose either $F$, $m$, or $\mu_k$ as the unknown variable.
    \item Describe the situation and explain the known variables.
    \item Imply that the object is moving/sliding at constant or uniform speed.
    \item When stating the angle, explicitly state if it is upward or downward with respect to the horizontal, but do not write any negative signs.
    \item If the object is pulled by a rope or a chain or similar objects, state the angle as the rope/chain forms an angle of \underline{\hspace{0.7cm}} with respect to the horizon.
    
    \item Ask students to calculate the unknown variable, and specify the unit. For example, find the force F in units of Newtons.
    \item Specify significant figures required for the answer.
\end{enumerate}
\end{promptbox}

\begin{promptbox}{Prompt 4: Solution Generation}
Write concise student facing solutions for the first 5 problems. Each solution should include the following elements:

\begin{enumerate}
    \item Explain that due to the object moving at constant velocity along a flat surface, the acceleration is zero in both directions, so the total force on both horizontal and vertical directions must both add up to zero according to Newton's second law of motion.
    
    \item Set positive x in the direction of the object's motion and positive y pointing up in the vertical direction.
    
    \item Write down the sum of forces along the $y$-axis equal to zero expression, starting with $\sum(\vec{F_y}) = \cdots = 0$.
        
    \item Similarly, write down the sum of forces along the $x$ axis equal to zero expression, using $f$ to represent friction.
    
    \item Find the friction force using the kinetic friction force equation, and find the normal force using the $y$-axis forces equation.
        
    \item Find the unknown variable.
    
    \item If the force is pointing downward, add a sentence clarifying that since the force is pointing below the $x$-axis, the numerical value of the variable $\theta$ should be negative the value used in the problem.
    
\end{enumerate}

Do not use numbers, instead, use languages similar to ``First, notice that...", ``Next, we can define...", ``We can then write down...".  
\end{promptbox}

\begin{promptbox}{Prompt 5: Formatting Prompt}
The last prompt asks the AI to output the generated problems in pre-determined yaml data format, and is omitted here.
\end{promptbox}

\begin{table}[!htbp]
\centering
\begin{subtable}[t]{0.32\textwidth}
\centering
\small
\begin{tabular}{lccc}
\toprule
\textbf{Scale} & \textbf{N} & NUM & MCQ \\
\midrule
\textbf{<4B} & \textbf{4} & 0.156 & 0.259 \\
\textbf{4-8B} & \textbf{6} & 0.534 & 0.536 \\
\textbf{14-32B} & \textbf{7} & 0.839 & 0.720 \\
\bottomrule
\end{tabular}
\caption{By Scale}
\label{tab:model_a}
\end{subtable}
\hfill
\begin{subtable}[t]{0.32\textwidth}
\centering
\small
\begin{tabular}{lccc}
\toprule
\textbf{Family} & \textbf{N} & NUM & MCQ \\
\midrule
\textbf{Qwen3} & \textbf{9} & 0.645 & 0.633 \\
\textbf{Llama3} & \textbf{3} & 0.167 & 0.219 \\
\textbf{Phi-4} & \textbf{4} & 0.628 & 0.562 \\
\textbf{GPT-oss} & \textbf{1} & 0.885 & 0.684 \\
\bottomrule
\end{tabular}
\caption{By Family}
\label{tab:model_b}
\end{subtable}
\hfill
\begin{subtable}[t]{0.32\textwidth}
\centering
\small
\begin{tabular}{lcc}
\toprule
\textbf{Variant} & NUM & MCQ \\
\midrule
\textbf{Base} & 0.336 & 0.590 \\
\textbf{Instruct} & 0.812 & 0.821 \\
\textbf{Thinking} & 0.915 & 0.795 \\
\bottomrule
\end{tabular}
\caption{By Variant (Qwen3-4B)}
\label{tab:model_c}
\end{subtable}
\caption{Model performance analysis across scale, family, and variant}
\label{tab:model_analysis}
\end{table}

\section{Impact of LM Scale and Architecture}
\label{app:b}

\subsubsection{Model Scale Effects.}

As shown in Table~\ref{tab:model_a}, model performance generally improves with model size. Small models ($<4$B parameters) achieve relatively low accuracy on both numerical-response and multiple-choice questions (NUM: 0.156, MCQ: 0.259), whereas larger models ($14$-$32$B parameters) achieve substantially higher accuracy (NUM: 0.839, MCQ: 0.720). This pattern suggests that smaller models may lack sufficient reasoning capacity for solving physics problems reliably. In contrast, larger models may achieve near-perfect performance, which can obscure the difficulty differences that are most relevant for modeling student performance. This result suggests that selecting models based on the expected difficulty of the questions and the target student population may improve the usefulness of LM-based validation for isomorphic problem banks.

\subsubsection{Architecture Family Differences.}

Table~\ref{tab:model_b} shows substantial variation across model families. GPT-oss-20B achieves high accuracy on numerical-response questions (0.885) but lower accuracy on multiple-choice questions (0.684), suggesting that its reasoning strengths may not transfer uniformly across answer formats. The Qwen3 and Phi-4 families show more balanced performance across formats, which may make them more reliable choices for cross-format validation. By contrast, the Llama-3 models generally underperform relative to the other model families, indicating possible limitations in physics reasoning.

\subsubsection{Training Variant Impact.}

The analysis of Qwen3-4B variants (Table~\ref{tab:model_c}) shows that instruction-tuned and reasoning-focused training improve physics problem-solving ability. The base model performs within the expected range for models of comparable scale, whereas the instruct-tuned model substantially improves accuracy across both numerical-response and multiple-choice formats. The thinking variant achieves very high numerical accuracy but shows a slight decrease in multiple-choice performance, possibly because extended reasoning can make the model more susceptible to over-analyzing distractors.

\subsubsection{Practical Implications.}

For isomorphic problem-bank validation, these findings suggest that mid-to-large-sized (4-32B parameters) instruct-tuned or reasoning-enabled models may serve as the most useful proxies. Models in this range operate at an accuracy level that better overlaps with the typical student performance distribution, as shown in Table~\ref{tab:results}, and are therefore more likely to reveal meaningful variation among problem variants. In contrast, models that are either too weak or too strong may be less effective for detecting difficulty outliers because of floor or ceiling effects.

\end{document}